\documentclass{article}

\usepackage[numbers]{natbib}
\usepackage{PRIMEarxiv}
\usepackage{multirow}
\usepackage[utf8]{inputenc} 
\usepackage[T1]{fontenc}    
\usepackage{hyperref}       
\usepackage{url}            
\usepackage{booktabs}       
\usepackage{amsfonts}       
\usepackage{nicefrac}       
\usepackage{microtype}      
\usepackage{lipsum}
\usepackage{fancyhdr}       
\usepackage{graphicx}       
\graphicspath{{media/}}     

\title{MoleNetwork: A tool for the generation of  synthetic optical network topologies}

\author{
  Alfonso~S\'anchez-Maci\'an, 
  Nataliia~Koneva, \\
  \textbf{Farhad~Arpanaei,
  José~Alberto~Hernández}\\
  Universidad Carlos III de Madrid \\
  Leganés \\
  Spain \\
  \texttt{alfonsan@it.uc3m.es} \\
   \And
  Marco~Quagliotti,
  Emilio Riccardi\\
  Telecom Italia \\
  Torino \\
  Italy \\
   \AND
  José~M.~Rivas-Moscoso, Juan~P.~Fernández-Palacios \\
  Telefónica I+D \\
  Madrid\\
  Spain
   \AND
  Li Zhang \\
  DTU Electro\\
  Technical University of Denmark (DTU)\\
  Kgs. Lyngby\\
  Denmark
}

\begin{document}
\maketitle

\begin{abstract}
Model networks and their underlying topologies have been used as a reference for techno-economic studies for several decades. Existing reference topologies for optical networks may cover different network segments such as backbone, metro core, metro aggregation, access and/or data center. While telco operators work on the optimization of their own existing deployed optical networks, the availability of different topologies is useful for researchers and technology developers to test their solutions in a variety of scenarios and validate the performance in terms of energy efficiency or cost reduction. This paper presents an open-source tool, MoleNetwork, to generate graphs inspired by real network topologies of telecommunication operators that can be used as benchmarks for techno-economic studies.
\end{abstract}

\keywords{Optical Network Topologies \and synthetic generation \and network architectures}

\section{Introduction}
Future generation services are expected to have very high demanding requirements in terms of bandwidth and delay for fixed and mobile networks. As an example, ITU-T Focus Group NET-2030 introduced several use cases for next-generation services \cite{ITU20} including Holographic Type Communication, Tactile Internet for remote operations, Intelligent operation networking, Network and Computing convergence, Digital Twins, Space-terrestrial integrated network, and Industrial Internet of Things with cloudification. They were later extended with five additional ones \cite{ITU20_2}. These use cases have different requirements in terms of bandwidth (e.g. capacity, QoS), time (e.g. latency, synchronization), security (e.g. privacy, trustworthiness), artificial intelligence (data computation, storage) or ManyNets (e.g. addressing, mobility).  

Providing resources for these increasing demands implies having a network planning strategy considering different scenarios depending on the expected market penetration of the services throughout a period of time. Network operators can use their internal topologies and models to forecast the required investments to support these services. Researchers and technology providers have to verify the advantages of their solutions (architectural or technical) under different conditions and topologies (e.g. \cite{1353397}).

Network topologies have been used as models for resiliency evaluation and techno-economic studies for some decades (e.g. \cite{MilanNet}, \cite{TokyoTopology}, \cite{Rueda2017RobustnessComparison}). Depending on the goal, they may cover different segments of the network, including backbone, metro core, metro aggregation, access and/or data center. For most of these segments, they are usually based on ring-based or mesh-based strategies \cite{arijs2000design, Farhad_6DMAN}.

In this work, topology means the physical layout of the network, i.e. the adjacencies between the nodes equipped with fiber. Fiber adjacencies allow the creation of optical line systems (OLS). In state of the art OLSs typically, but not exclusively, rely on Dense Wavelength Division Multiplexing (DWDM) systems in C-band only. In the future it is expected that multiband systems will be introduced (C+L systems are actually already commercial products) or high capacity systems that use special fibers such as multi core or hollow core fibers. Nodes are physical places where equipment of different network layers can be installed, such as optical equipment (e.g, ROADM), OTN switches, packet layer switches (e.g., IP routers) and also servers that implement virtual telco network and service functions. The topology of higher logical layers, such as OTN or IP, implies further assumptions in a multilayer perspective that is not considered in this article. The nodes in the topological models presented in this article are therefore the places where the multiplexing/demultiplexing and optical switching functionalities are carried out, at a minimum.

This paper presents MoleNetwork, an open-source tool to generate reference backbone and metro physical topologies to be used as a basis for techno-economic studies. The rest of the paper is organised as follows. Section \ref{background} presents some background and state-of-the-art in terms of topologies, structures, and methodologies to generate them. Section \ref{statistics} describes part of the statistical information provided by operators in the context of some European projects.
Section \ref{solution} introduces the tool and the different parameters and possibilities to use it and presents an example of a scenario implementing a topology from the backbone to the metro aggregation sections. Section \ref{validation} compares the output statistics from the tool with the corresponding inputs to check that it is able to provide the desired results. Finally, Section \ref{conclusion} presents the conclusions.

\section{Background}
\label{background}

There have been several efforts in the last decades to provide topologies for research and innovation studies. 

Real topologies (or an approximation to them) have been published in several studies, mainly for backbone networks (e.g. Italy \cite{Italia_topo}, Germany \cite{Germany_topo} or Portugal \cite{Portugal_topo}) or research networks (e.g. \cite{rediris}).
Many of them were collected in the Internet Topology Zoo \cite{Zoo}, last updated in 2013.

The generation of realistic synthetic topologies has also been an interesting field of study. Numerous studies have focused on generating topologies that mimic the structure of the Internet e.g. \cite{nem}, \cite{realnet}. In \cite{waxman}, random graph generation is completed by allocating nodes in a plane and creating links between two nodes based on a probability distribution that considers the Euclidean distance between them. These models typically adhere to a power-law degree distribution, similar to that observed in the Internet \cite{powerlaw} and other scale-free networks. 

Optical transport networks with survivable topologies differ from scale-free networks \cite{pavan} in several characteristics, making these Internet-based models inaccurate. Specific research has been carried out for these types of networks. In \cite{pavan}, a method is proposed to generate realistic optical transport topologies with an average nodal degree within a particular range. The total area of the plane is divided into regions, and the minimum distance between nodes and the total number of nodes, which are input parameters, are used to allocate nodes within the regions. Connections between nodes (within regions and between regions) in the topology follow specific rules to ensure survivability, delegating to Waxman method \cite{waxman} to complete the network connections. Authors in \cite{TokyoTopology} define a methodology based on the information of regional railways to create the geometric structure of the Metropolitan Area Network (MAN) topology as there is a strong correlation with the population densities. They create a MAN model with 23 nodes for Tokyo (Tokyo23) and a compression algorithm to reduce the number of nodes which results in another topology Tokyo12. They compare the results with another model design using communication network information.

Some papers also focused on generating 5G infrastructures. 5GEN \cite{5GEN} starts from Active Antenna Units (AAUs) and creates clusters of 6 AAUs based on their location (or population density \cite{5GEN2}) connecting them to a switch. Then, it creates access and aggregation rings grouping switches from different hierarchical levels. INFRA5YNTH \cite{INFRA5YNTH} defines random populations and deploys Radio Access Networks nodes to serve a number of cells following existing specifications and grouping traffic in access rings and aggregation rings. Both approaches are more oriented to a greenfield scenario.

In this paper, we present MoleNetwork, a tool that is capable of generating Backbone, Metro core and Metro aggregation structures based on the topology information provided by operators in the context of the EU-funded ALLEGRO project\footnote{ALLEGRO 'Agile ultra-low energy secure networks', \url{https://www.allegro-he.eu/}, last access March 2024}. This tool provides several options to generate topologies including different structures for metro core (mesh or n-ring structure), the possibility of defining the connectivity degree distribution for mesh structures, the ability to create a backbone to metro aggregation integrated mode or the specification of the different types of nodes for the hierarchy and their distribution within a level, among others. Depending on their role in the network hierarchy, nodes can be national, regional or local. They normally perform the add and drop of locally terminated traffic and switch both local traffic and transit traffic. Some nodes may not be enabled to handle add and drop traffic and they perform only the transit function, in that case they are classified as pass-through (or pure transit) nodes.

\section{Topology data}
\label{statistics}
The implemented MoleNetwork tool takes advantage of statistical and topological information defined within the ALLEGRO European-funded project. Operators taking part in the project defined some guidelines for backbone, metro core and metro aggregation structures to perform techno-economic studies covering increased demands in terms of capacity and latency due to next-generation services being deployed into the networks \cite{Farhad_6DMAN}.

\begin{figure}[ht]
  \centering
  \includegraphics[width=250pt]{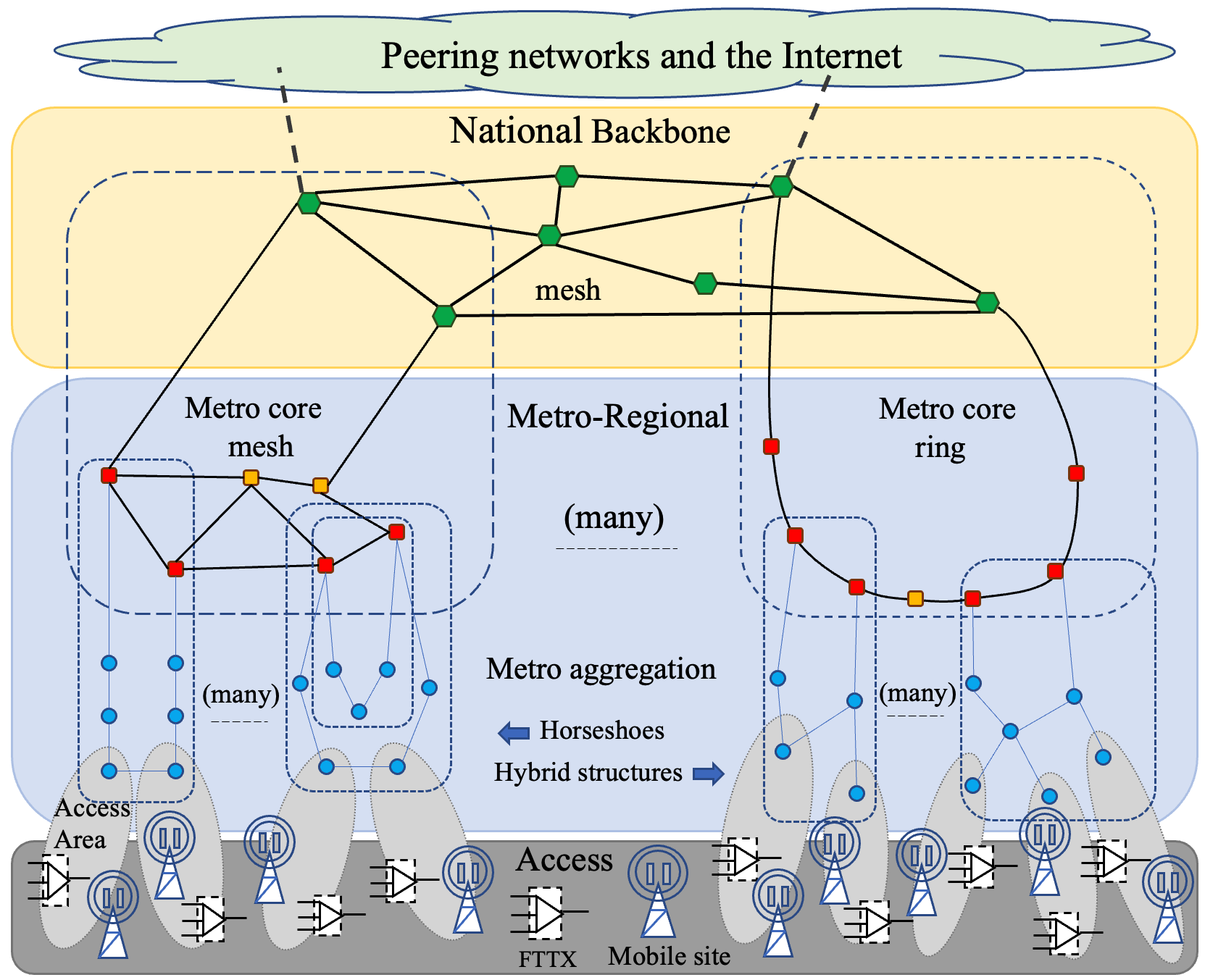}
  \caption{\small Typical hierarchical physical topology of a network operator.}
  \label{Hierarchy_Sample}
\end{figure}

Figure \ref{Hierarchy_Sample} gives an overall view of an operator's physical network, from the access to interconnections with other networks and with the Internet, highlighting the optical layout topologies typical of each segment. Starting from the bottom, the access collects the FTTX-type fiber bindings (fiber to the home, curb, antenna etc..) and the direct fiber links of the radio mobile sites and business customers. Access is divided in areas. Each area (gray ellipses in figure \ref{Hierarchy_Sample}) is collected by a local node. This network segment (the access) is not considered in the article and is the subject of future developments of MoleNetwork. Above the access, Figure \ref{Hierarchy_Sample} shows the metro-regional segment which, in this particular but highly representative example, is divided into two segments: an aggregation segment, which collects the access devices (e.g., PON OLTs, switches aggregating mobile traffic), and a core segment. The aggregation segment is typically composed of double-ended horseshoes, which guarantee good survivability in case of single link or node failures, as can be seen in the left part of the figure. Other structures such as trees, meshes, or hybrid topologies, imposed by infrastructural or economical constraints (e.g., it is not always possible to deploy fiber to make horseshoes or rings), are also possible, as shown on the right. The metro aggregation structures are connected to the metro core network which, as shown in the central band of the figure, can be a mesh (on the left) or a ring (on the right). The third and higher network segment is the national backbone represented in the upper part of the figure. In this case, the typical topology is a mesh and some variants are possible such as the presence of twin nodes and parallelized topology which guarantee a high level of geographic survivability. Finally, above the national backbone, the external networks are represented, including the Internet: they are connected to the backbone with specific few interconnection links.

Information about two backbone topology structures were defined by the operators. Both of them are implemented as a mesh network with multiple nodes. 

In the first case, the backbone is composed of national nodes (mainly), some regional nodes, and some transit/pass-through nodes. With about 50 nodes and 80 links, the statistics about the topological degree (connectivity of the nodes) and the link lengths are defined in tables \ref{tab:topo_degree1} and \ref{tab:topo_link1}. 

\begin{table}[ht]
\centering
\caption{Topological degree statistics of first backbone structure}
    \begin{tabular}{|c|c|}
    \hline
        Topological Degree & Occurrence \\
    \hline
        1 & 0 \\
    \hline
        2 & 22.7\% \\
    \hline
        3 & 40.9\% \\
    \hline
        4 & 27.3\% \\
    \hline
        5 & 9.1\% \\
    \hline
        6 & 0 \\      
    \hline
    \end{tabular}
\label{tab:topo_degree1}
\end{table}

\begin{table}[ht]
\centering
\caption{Link length statistics of first backbone structure}
    \begin{tabular}{|c|c|}
    \hline
        Link length range (kms) & Occurrence \\
    \hline
        0-50 & 15.5\% \\
    \hline
        50-100 & 16.9\% \\
    \hline
        100-200 & 33.8\% \\
    \hline
        200-400 & 25.4\% \\
    \hline
        400-600 & 8.5\% \\     
    \hline
    \end{tabular}
\label{tab:topo_link1}
\end{table}
The other structure includes distributing two nodes per region, with some exceptions in highly populated regions which would require a higher number of nodes. Nodes are divided into hierarchical levels (L1, L2 and L3) with pass-through nodes also included. With more than 125 nodes and about 190 links, statistics about nodal degree connectivity are shown in table \ref{tab:topo_degree2}.

\begin{table}[ht]
\centering
\caption{Topological degree statistics from the second backbone structure}
    \begin{tabular}{|c|c|c|c|c|}
    \hline
       \multirow{2}{*}{ Hierarchical Level} & \multirow{2}{*}{\% of nodes} & \multicolumn{3}{c|}{ {  Connectivity degree}} \\
    \cline{3-5}
         &  & Avg. & Min. & Max \\
    \hline
        L1 & 4\% & 6 & 3 & 9\\
    \hline
        L2 & 12.6\% & 4.8 & 3 & 9\\
    \hline
        L3 & 77.8\% & 3.4 & 3 & 9\\
    \hline
        Pass-through & 5.6\% & 3.6 & 2 & 4 \\    
    \hline
    \end{tabular}
\label{tab:topo_degree2}
\end{table}

\begin{figure}[ht]
  \centering
  \includegraphics[width=250pt]{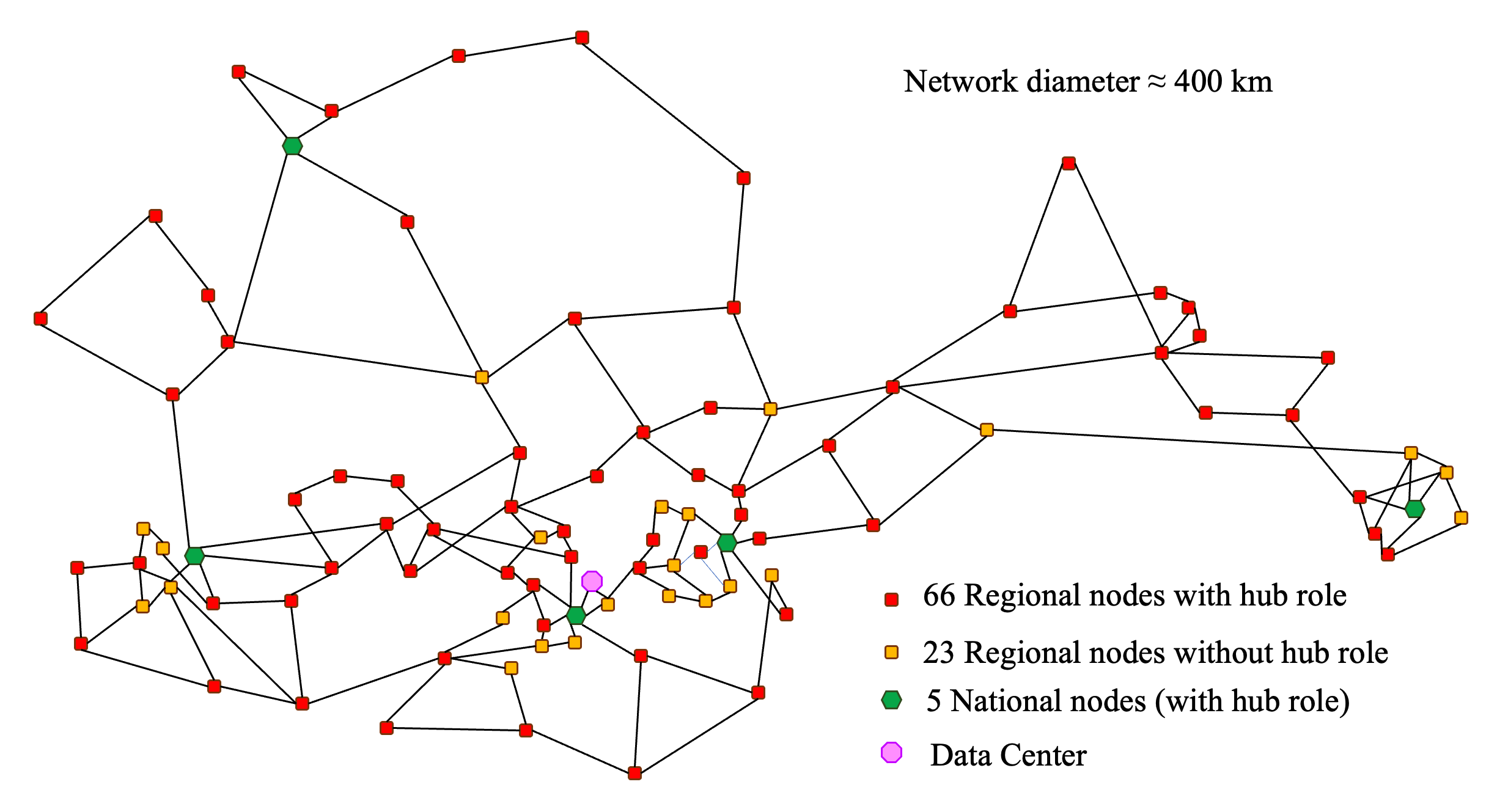}
  \caption{\small Sample of a large size  metro-regional meshed core network.}
  \label{MetroMesh_Sample}
\end{figure}

For the regional metro core structure, there are again two different structures. One is, again, based on mesh networks and it can include a variable number of nodes depending on a specific metro region. An example of a metro regional mesh for a big macro-region is included in Figure \ref{MetroMesh_Sample}. In this sample network, whose network diameter is approximately 400 km, the majority are regional nodes (red squares) that play the role of a hub for aggregation structures (double-hubbed horseshoes). There are other regional nodes that take part of the core mesh even if they do not act as a hub for any of the aggregation horseshoe (orange squares). Five national nodes (green hexagons) take part of the metro core mesh assuring the interconnection with the backbone and there is also a special node hosting a big Data Center (polygon in violet). Regional metro networks smaller than the one illustrated in Figure \ref{MetroMesh_Sample}, with a few dozen of nodes, or even much smaller (around twenty nodes), with network diameters in the range of 100-300 km, are typical of regions less extensive than that of the example shown in the figure.

Statistics about link lengths and types of nodes in the region are included in Table \ref{tab:topo_metro_link1} and \ref{tab:topo_metro_type1}.

\begin{table}[ht]
\centering
\caption{Link length statistics of first metro core structure}
    \begin{tabular}{|c|c|}
    \hline
        Link length range (kms) & Occurrence \\
    \hline
        0-10 & 39\% \\
    \hline
        10-40 & 37\% \\
    \hline
        40-80 & 21\% \\
    \hline
        80-120 & 3\% \\ 
    \hline
    \end{tabular}
\label{tab:topo_metro_link1}
\end{table}

\begin{table}[ht]
\centering
\caption{Node type statistics of first metro core structure}
    \begin{tabular}{|c|c|}
    \hline
        Node type & Occurrence \\
    \hline
        Data centre & 1\% \\
    \hline
        National Central Office & 5\% \\
    \hline
        Regional Central Office & 70\% \\
    \hline
        Regional Central Office (no hub) & 24\% \\ 
    \hline
    \end{tabular}
\label{tab:topo_metro_type1}
\end{table}

\begin{figure}[ht]
  \centering
  \includegraphics[width=250pt]{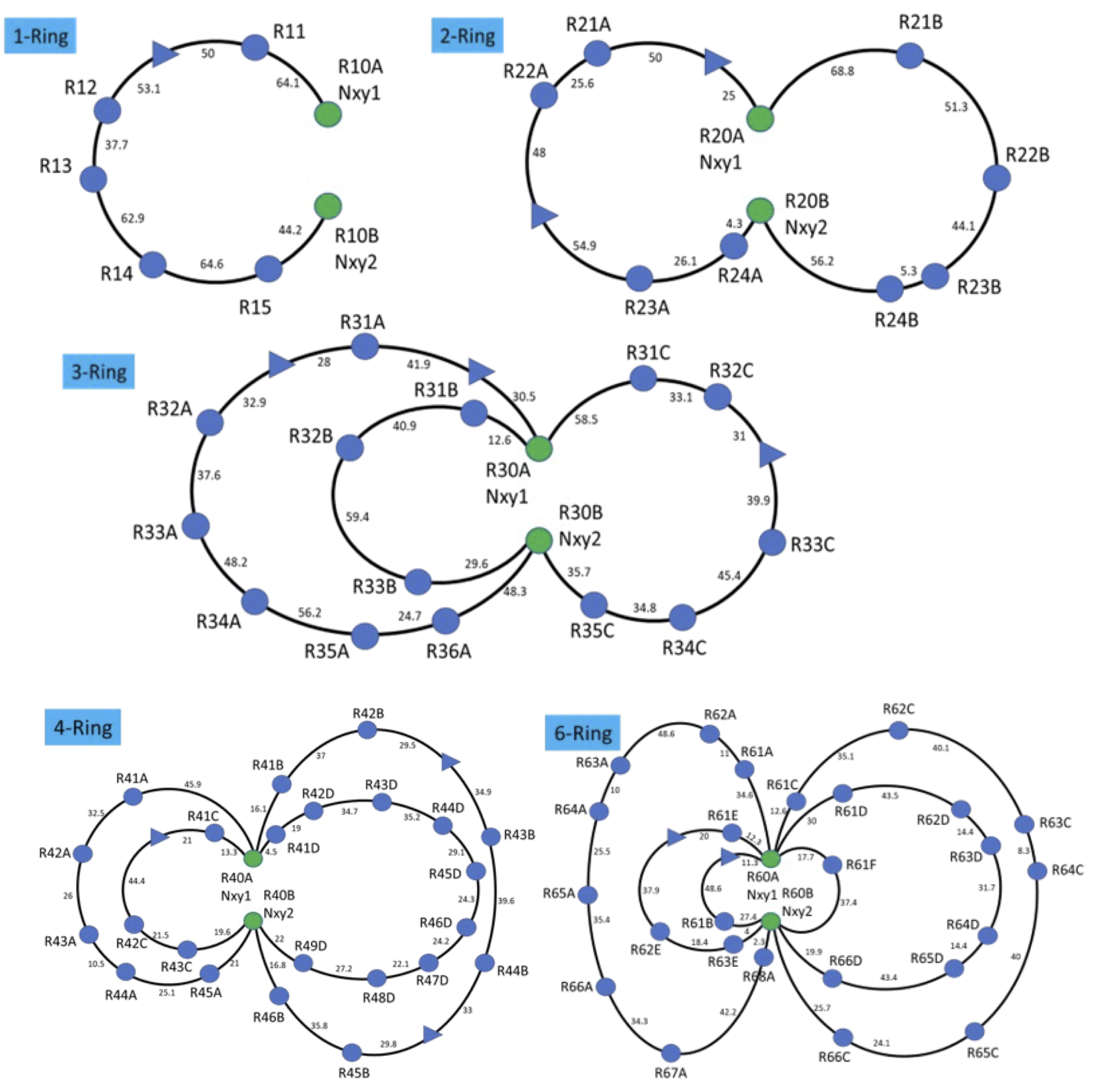}
  \caption{\small Examples of regional metro core  ring structures (in green national nodes, in blue regional nodes).}
  \label{ring_structures}
\end{figure}

The second design for the regional metro core structure is based on ring structures (Figure \ref{ring_structures}). Green nodes are national nodes terminating the metro rings, blue nodes are regional nodes taking part of the rings and used as hubs for metro aggregation structures, while triangles represent amplification sites.

In particular, structures are defined with 1, 2, 3, 4, and 6 rings with different probabilities as shown in Table \ref{tab:topo_metro_ring1}. For each of these structures, several parameters can be enumerated including the average length of each ring in the structure and the average number of offices per ring.

\begin{table}[ht]
\centering
\caption{Statistics for the occurrence of n-ring structures}
    \begin{tabular}{|c|c|}
    \hline
        Structure & Occurrence \\
    \hline
        1-ring & 8\% \\
    \hline
        2-ring & 53\% \\
    \hline
        3-ring & 25\% \\
    \hline
        4-ring & 10\% \\
    \hline        
        6-ring & 4\% \\
    \hline  
    \end{tabular}
\label{tab:topo_metro_ring1}
\end{table}

\begin{figure}[ht]
  \centering
  \includegraphics[width=250pt]{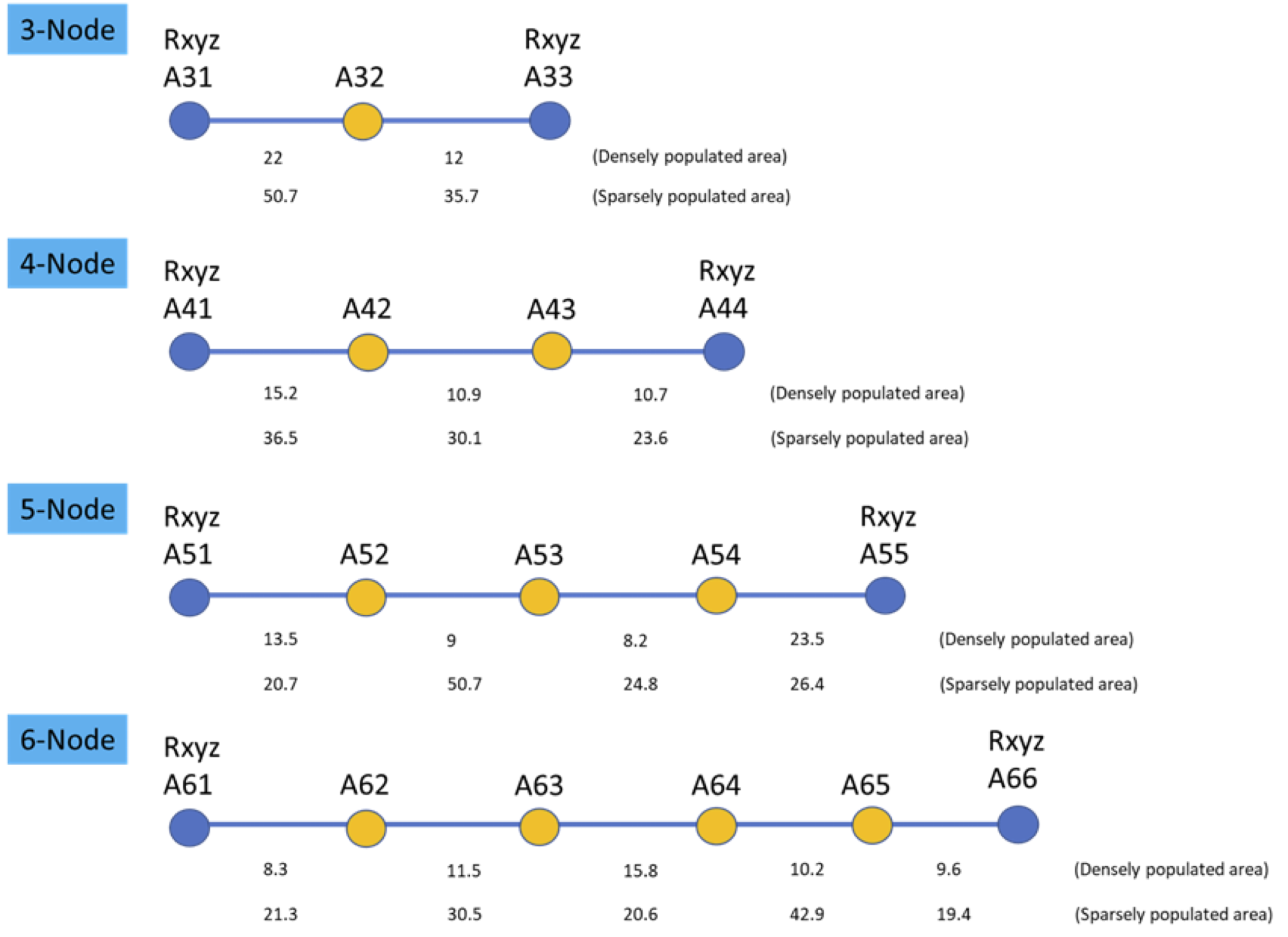}
  \caption{\small Examples of regional metro aggregation horseshoes (in blue regional end nodes, in yellow local aggregation nodes).}
  \label{ring_aggregation}
\end{figure}

Finally, the metro aggregation networks are designed based on a horseshoe structure where its ends are nodes from the regional metro core structure. The number of nodes composing the horseshoe may vary from 3 to 6 (including both regional end nodes - in blue - and local nodes - in yellow - as shown in Figure \ref{ring_aggregation}), with one of the operators having up to 9 nodes in some of them. Statistics for the number of nodes and the metro aggregation design from one of these operators are shown in Tables \ref{tab:topo_agg_ring1} and \ref{tab:topo_agg_ring2}.

\begin{table}[ht]
\centering
\caption{Statistics for the number of nodes in a metro aggregation horseshoe}
    \begin{tabular}{|c|c|}
    \hline
        Number of nodes & Occurrence \\
    \hline
        3 & 10\% \\
    \hline
        4 & 19\% \\
    \hline
        5 & 21\% \\
    \hline
        6 & 27\% \\
    \hline        
        7 & 14\% \\
    \hline
        8 & 5\% \\
    \hline
        9 & 4\% \\
    \hline
    \end{tabular}
\label{tab:topo_agg_ring1}
\end{table}

\begin{table}[ht]
\centering
\caption{Statistics of the metro aggregation horseshoe}
    \begin{tabular}{|c|c|c|c|c|}
    \hline
        & min & max & average & std. deviation \\
    \hline
        Num hops (hub-to-hub) & 2 & 8 &4.5 & 1.5 \\
    \hline
        Link length (km) & 0.6 & 166.4 & 21.4 & 22.2 \\
    \hline
        Total horseshoe length (km) & 15.2 & 301.9 & 95.6 & 61.7\\

    \hline
    \end{tabular}
\label{tab:topo_agg_ring2}
\end{table}

This information is useful in order to create different topologies that represent realistic scenarios for techno-economic studies. They are the basis for the design of our proposed solution, i.e. MoleNetwork tool.

\section{Proposed solution}
\label{solution}
The MoleNetwork tool is split into several modules that can be combined together. It follows a pluggable infrastructure so different strategies for backbone/metro network generation and clustering can be added. It is written in Python and is supported by widely used libraries such as Matplotlib, numpy, sklearn, scipy or Networkx among others. The tool is open source and it is publicly available\footnote{https://github.com/amacian/molenetgen} together with an online tutorial\footnote{https://www.youtube.com/watch?v=Fr0dzMEXKnM}.

\subsection{Backbone generation and clustering}
\begin{figure}[ht]
  \centering
  \includegraphics[width=250pt]{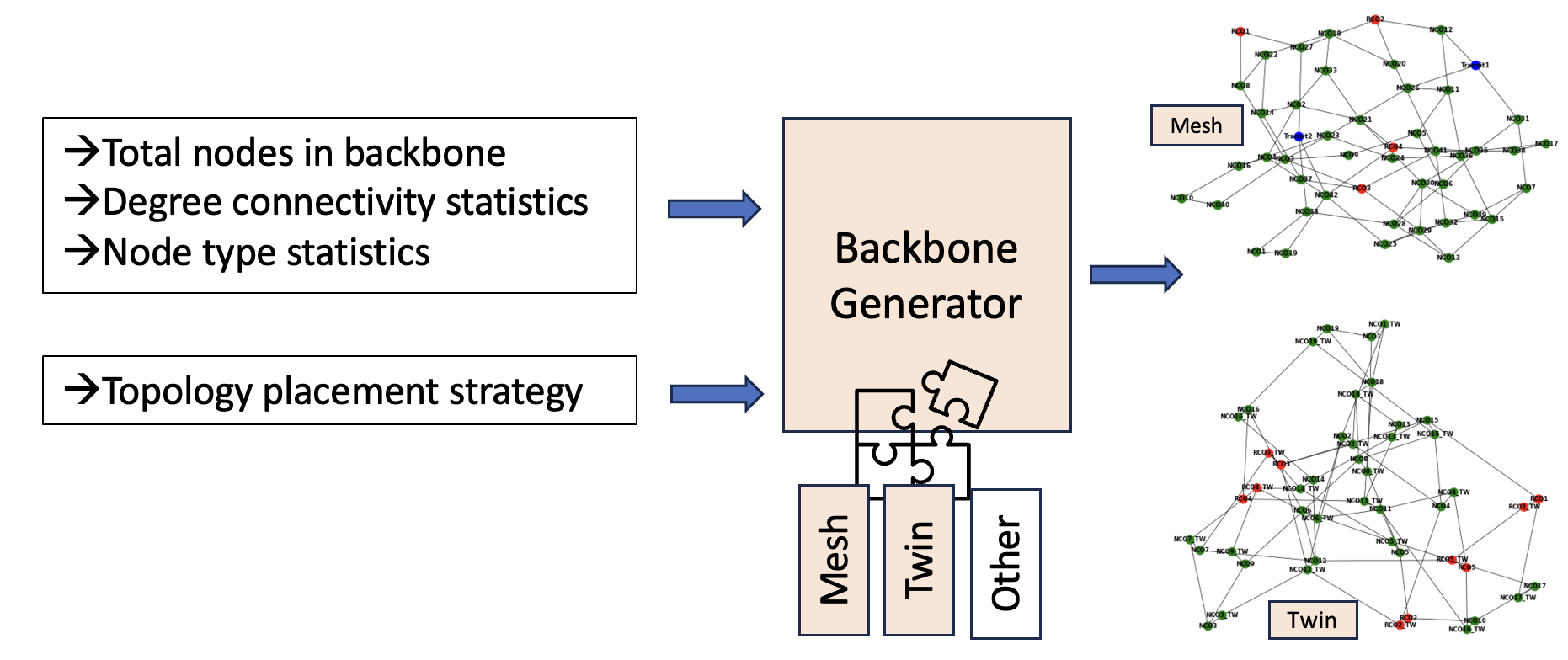}
  \caption{\small Backbone Generation process.}
  \label{BB_gen}
\end{figure}

Backbone generation (Figure \ref{BB_gen}) is provided using a generic abstract model that can be implemented following different strategies. The tool brings by default three possible options, 1) generating a mesh network based on statistical values of parameters, 2) creating a set of interconnected twin nodes near to each other and 3) using an approach similar to the one proposed in \cite{pavan}. Additional strategies can be added in a modular way.

For the mesh network strategy, the user provides statistics regarding degree connectivity such as the ones in Table \ref{tab:topo_degree1} ($degree$, $probability$), percentage of nodes $perc\_type$ per type (e.g. national or regional central offices) similar to the information of the columns Hierarchical level and percentage of nodes in Table \ref{tab:topo_degree2}, and the total number of nodes $nodes$ in the network. It also selects one of the available topology placement ($layout$) strategies (e.g. Spring or Kamada-Kawai\cite{Kamada}) from the underlying libraries.

The algorithm includes the following steps:
\begin{enumerate}
\item Get the degree connectivity statistics (i.e. $degree$, $probability$) and the total number of nodes ($nodes$).
\item Generate a random number $degree_i$ for the degree of connectivity of each node, i.e. the number of links to be created from each node $i$. This random number is based on the statistical information ($degree$ and $probability$) provided as input. If the sum of the total number of expected links ($\sum_{i=1}^{nodes}degree_i$)is not even, increase the value to 1 for one of the nodes with the lowest degree.
\item Generate random connections $edges$ between nodes based on the previous connectivity values using the configuration model \cite{confmodel1}\cite{confmodel2}.
\item Drop parallel edges and self-loops.
\item Validate that the graph is connected and check for node survivability. If fail, repeat steps 2 to 4.
\item Ensure edge survivability by applying k edge augmentation \cite{k_edge} if needed.
\item Get the input statistics about node types $perc\_type$, and assign types randomly to the created nodes.
\item Rename the nodes based on their type and sequence. e.g. NCO14.
\item Assign coordinates (positions $pos$) based on the selected layout strategy.
\item Define actual distances $dists$ by scaling up the existing layout using the defined distance ranges (and, in particular, the maximum distance).
\item Assign colors for the graph, based on the assigned types.
\end{enumerate}

The twin nodes strategy uses the same input parameters, just forcing the total number of nodes to be even. The process will place half of the nodes with a reduced topology including all links and then create the twin nodes next to the original ones and split the original links between the twins. This is detailed in the following steps:
\begin{enumerate}
\item Get the degree connectivity statistics (i.e. $degree$, $probability$) and the total number of nodes ($nodes$).
\item Choose only half of the nodes adapting the degree connectivity statistics so $degree_{new} = (degree_{old}*2)-2$. This will ensure, as explained later, that the final connectivity will match the original one as the original and twin nodes will be connected to each other.
\item Follow steps 2 to 9 from the mesh topology algorithm. 
\item Generate a twin node $node\_tw_i$ for each node (except transit nodes) by placing it next to the original one $node_i$ within a reduced fixed distance range $d_{range}$. The actual coordinates of the twin node $x\_tw_i, y\_tw_i$ are generated by defining the distance in the coordinates $x$ and $y$ in relation to $node_i$ based on random values within $d_{range}$. 
\item Relabel the new nodes by adding a specific suffix ($\_TW$) to the original corresponding one (e.g. $NCO1$ and the twin $NCO1_{TW}$.
\item Copy the assigned node type from the original node to the twin one. 
\item Split randomly in two equal size sets ($edgesA_i$ and ($edgesB_i$) the connections ($edges_i$) that were created between the original node $node_i$ and third nodes. One of the subsets (e.g. $edgesA$) is reassigned to the new node $node\_tw_i$, creating the connections from this twin node to the third nodes and dropping the original connections that existed from $node_i$ to them.
\item Create the link between the original and twin nodes.
\item Define actual distances $dists$ by scaling up the existing layout using the defined distance ranges (and, in particular, the maximum distance).
\item Assign colors for the graph, based on the assigned types.
\end{enumerate}

The strategy based on \cite{pavan} uses the average degree connectivity instead of the whole distribution to build a network. As previously explained, the nodes are allocated in regions and some rules are applied to the connections within a region and between regions, using also Waxman \cite{waxman} probability to complete the link distribution. The original paper provides a more detailed description.

\begin{figure}[ht]
  \centering
  \includegraphics[width=250pt]{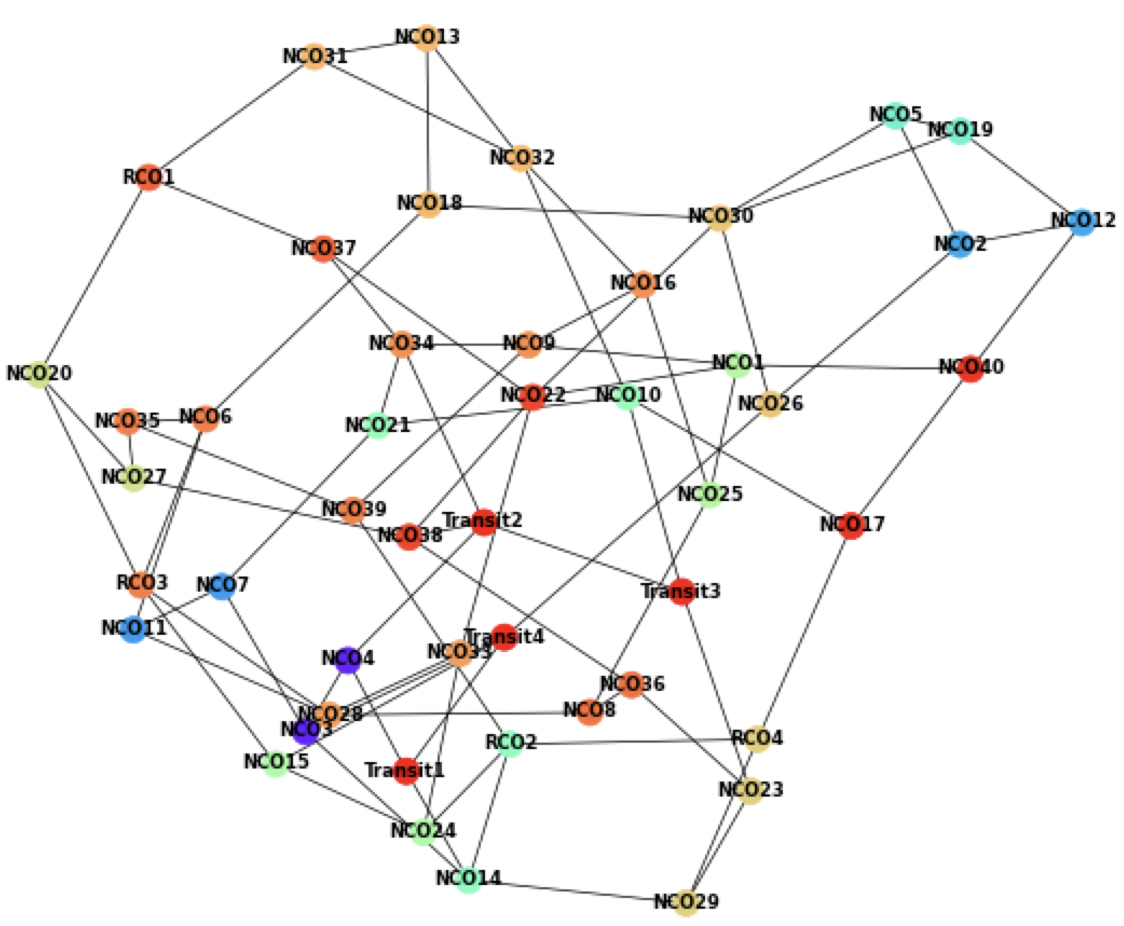}
  \caption{\small Examples of clustering for metro region creation.}
  \label{Cluster}
\end{figure}

An additional functional component is available in order to cluster the backbone nodes to create the metro regions for later metro core network generation. Two different implementations have been completed, i.e. clustering based on only the distance between nodes, and another option taking into account the distance and connectivity. Both of them use a "Density-based spatial clustering of applications with noise" (DBSCAN \cite{DBSCAN}) clustering algorithm where the user can provide the $\epsilon$ parameter (neighborhood size) and also they can force to avoid leaving single nodes $avoid\_single$ (pairing them with existing ones). An example of the result of the process of creation of metro regions is included in Figure \ref{Cluster}.

The detailed algorithm for the distance-only based clustering is as follows:
\begin{enumerate}
\item Select the nodes $nodes$ from the topology and exclude those that are  transit only nodes.
\item Get the coordinates $coord$ of all the selected nodes and apply the DBSCAN algorithm using the provided $\epsilon$ as the parameter, selecting the Euclidean distance as the one to be used for clustering.
\item For each node $node_i$, retrieve its cluster label $label_j$ (corresponding to the cluster $j$). These clusters may include just one node or several of them depending on the topology and the $\epsilon$ parameter.
\item Group all the transit nodes into a virtual cluster with a new label $label_{max}$.
\item If single nodes are not allowed (input parameter $avoid\_single$ is true), continue to the next step, otherwise, end the process.
\item Go through the cluster nodes $label_j$ selecting the ones that include just one element $node_i$ and use a K-Dimensional tree to find the nearest neighbor from the node $node_i$ among the rest of the nodes. This $node_{ngh}$, included in another cluster $label_{ngh}$, could be another single-clustered node, or, alternatively, $label_{ngh}$ may have several members.
\item Place the node $node_i$ together with the nearest neighbor $node_{ngh}$ in the neighbor's cluster $label_{ngh}$.
\end{enumerate}

For the distance and connectivity-based clustering, the previous process is specialized so only the elements that are connected to each node are considered in order to build the cluster, i.e. two nodes $node_i$ and $node_j$ can belong to the same cluster only if there is an edge $edge_ij$ between them or they are connected through other nodes from the same cluster (e.g. $edge_ik$ and $edge_kj$ with $node_k$ also belonging to the cluster). This is done for the original construction (separating those nodes that are not connected) and also for the reassignment of single nodes (considering only the nodes connected when finding $node_{ngh}$).

\begin{figure}[ht]
  \centering
  \includegraphics[width=250pt]{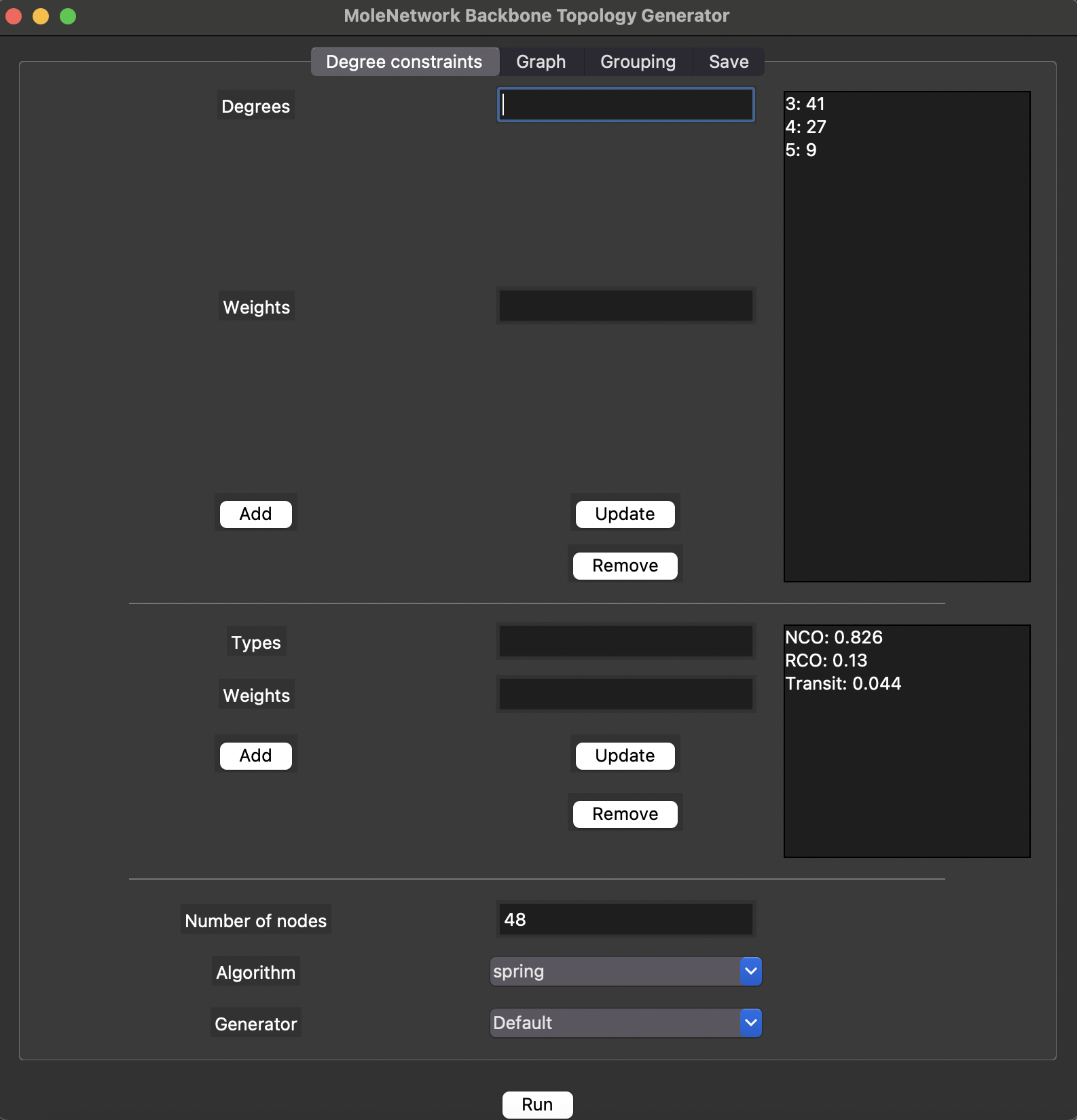}
  \caption{\small GUI screen for parameter definition.}
  \label{GUI_BB_1}
\end{figure}

\begin{figure}[ht]
  \centering
  \includegraphics[width=120pt]{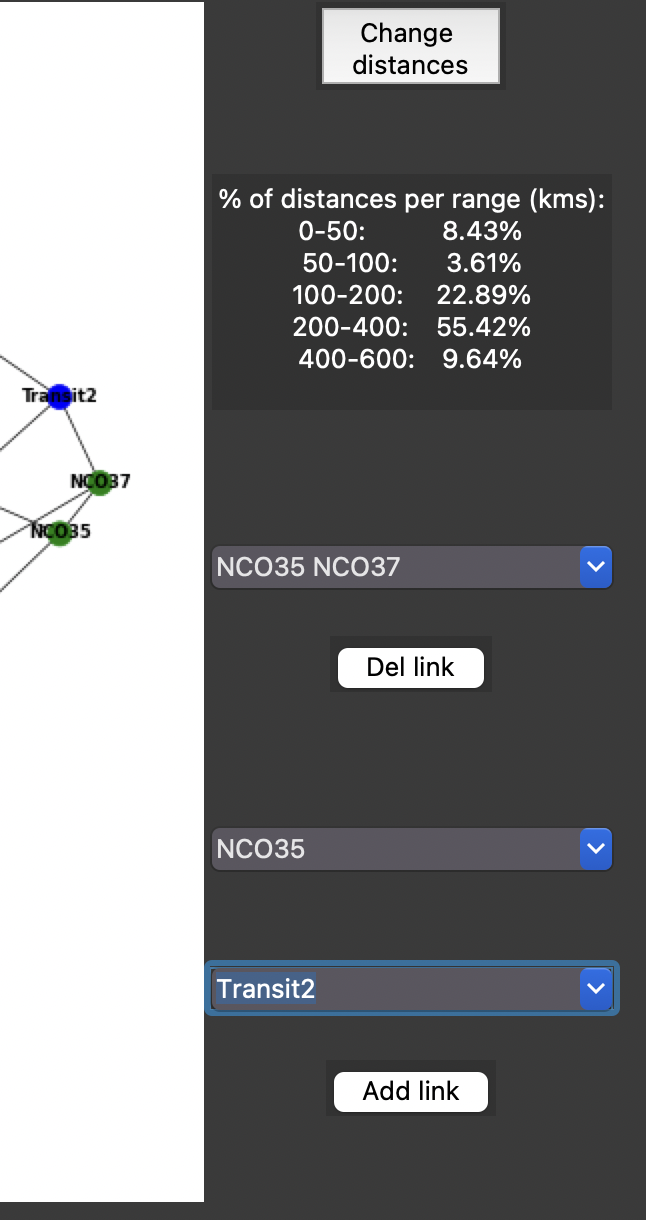}
  \caption{\small Additional functionalities through the GUI.}
  \label{GUI_BB_2}
\end{figure}

Taking into account that the user might want to customize the structure, a graphical user interface has been created (see Figure \ref{GUI_BB_1}). In this interface, the user can fine-tune the specific input parameters, choose the appropriate implementation of the backbone generator and clustering mechanisms (with the related constraints), but it can also create and drop links \ref{GUI_BB_2} from the generated architecture and also define link length ranges (setting the longest link distance) to get statistics from the different generated lengths.

Results (nodes, links, and additional information) are currently being written to an Excel file (and the topology to an image file) to be used as input for other tools.

Notice that this Excel file can then be manually modified or adapted to change any information as desired (e.g. modifying the metro regions or dropping nodes or links).

\subsection{Metro core generation}
Two options are provided in the tool to generate Metro core structures: generating a specific metro core network by defining only statistical parameters, or reading clusters created by the backbone module.

Metro generation is also driven by a generic model that is implemented through different strategies. In this case, it is possible to create a metro mesh or a ring-based topology for a metro region.

When creating a metro mesh structure, it requires similar parameters as in the backbone, i.e. degree connectivity statistics, number of nodes per type, the total number of nodes, and the algorithm for layout. The behavior of this part is similar to the mesh generation for backbone networks.

If information is read from the previously generated backbone network, it is possible to choose one of the created metro regions and build a mesh network based on it. In this case, the previously commented parameters are also considered. The only change has to do with the percentage of national central offices. This value is no longer used and, instead, the actual nodes defined within the cluster are utilized.

The algorithm for creating a metro core mesh structure is described next:

\begin{enumerate}
    \item The main nodes (national nodes or the ones defined in the metro cluster) are selected independently to generate a subnet.
    \item The number of nodes defined per node type is distributed evenly among all the subnets.
    \item For each of the defined subnets, a topology of nodes is created following the steps 2 to 7 of the original backbone generation algorithm, i.e. from random link number generation to type assignment.
    \item Each subnet is connected to the previous one forming a new topology by: \begin{enumerate}
        \item Renaming the nodes of the subnet to avoid conflicts and to follow a sequence.
        \item Choosing two random nodes from both subnets $s_{1a}$, $s_{1b}$ and $s_{2a}$, $s_{2b}$ among those with the minimum degree connectivity.
        \item Creating links between these nodes $s_{1a}$ to $s_{2a}$ and $s_{1b}$ to $s_{2b}$
    \end{enumerate} 
    \item Colors are assigned to the different nodes depending on their type.
    \item Coordinates are assigned based on the layout strategy that was chosen.
    \item Actual distances are defined by scaling up the layout as in previous algorithms.
\end{enumerate}

It is also possible to define a ring-based approach for metro core generation. For this, the particular implementation uses the definition of different ring-based structures and statistics made by Telef\'onica in the context of the European Allegro project and it ignores any of the parameters defined for metro mesh generation. 

These structures use two nodes from the backbone as the main nodes and one or several (1, 2, 3, 4 or 6) rings are deployed between them. The main inputs for this approach are the number of rings $nrings$ composing the structure (1-6) and the name of the two backbone nodes ($end1$ and $end2$) used as ends of the rings (e.g. NCO15 and NCO15\_TW). Optionally, a prefix ($pref$) and an initial index ($init\_idx$) can also be provided to customize the name of the newly generated nodes (e.g. to associate them to the cluster they belong or to the end nodes). Although average lengths are defined for the different segments of the rings, a variability parameter ($var$) is also optionally provided to introduce some noise in the actual length value for each segment of every ring. 

The creation of these ring structures is divided into several steps:
\begin{enumerate}
    \item An empty graph $ring$ is initially created.
    \item For each ring $r_i$ composing the N-ring structure (being $i$ between 1 and $nrings$), the following steps are taken:
    \begin{enumerate}
        \item A specific $suffix$ is assigned ("A" to "F") depending on the position of the ring within the structure (e.g. "A" to the first ring and "F" to the sixth if it exists). 
        \item The average total length $l_{i}$ of the ring $i$, the number of offices it includes $n_{i}$, the length range ($min\_length_{ij}, max\_length_{ij}$) for each segment $j(=n_{1}+1)$ of the ring $i$, and the maximum distance without amplifiers $distamp_{i}$ are retrieved from the initial configuration parameters. These values depend on the specific N-ring structure and the particular ring within it.
        \item The actual total length value of the ring is calculated based on $l_{i}$ and $var$, using a uniform distribution.
        \item The final values for the length of each segment $j$ of the ring $length_{ij}$ (except the last one) are calculated based on a uniform distribution between $min\_length_{ij}$ and $max\_length_{ij}$  
        \item The length of the last segment is calculated by subtracting $\sum_{j=1}^{n_{i}}length_{ij}$ from $l_{i}$. If the result is negative, it is weighted into a positive value, reducing the length of the rest of the segments to keep $l_{i}$ as the total length. 
        \item The segments are randomly reallocated within the link.
        \item The name of the offices is generated sequentially concatenating the prefix $pref$, the number of rings $nrings$ of the structure, an increasing index starting at $init\_idx$ for each ring, and $suffix$.
        \item Amplifiers are placed by breaking the links with distances higher than $distamp_{i}$. Names for the amplifiers are created in a similar way, but include the "AMP" string in the name.
        \item Nodes $end1$ and $end2$ are inserted at the beginning and end of the ring.
        \item A $subring$ graph is created with the information generated in the previous steps.
        \item The $subring$ is merged with the original $ring$.
    \end{enumerate}
    \item Graph distances are retrieved.
    \item Positions are generated by applying a spring layout to the $ring$ graph.
    \item Types are defined by assigning a "national" type to $end1$ and $end2$ and a "regional" type to the rest of the nodes.
    \item Colors are assigned to the nodes based on their type.
    \item For each node of each ring, one of the end nodes ($end1$ or $end2$) is selected as the reference national node based on the distance, using the shortest path algorithm and a dummy node connected to the ends.
\end{enumerate}

\begin{figure}[ht]
  \centering
  \includegraphics[width=250pt]{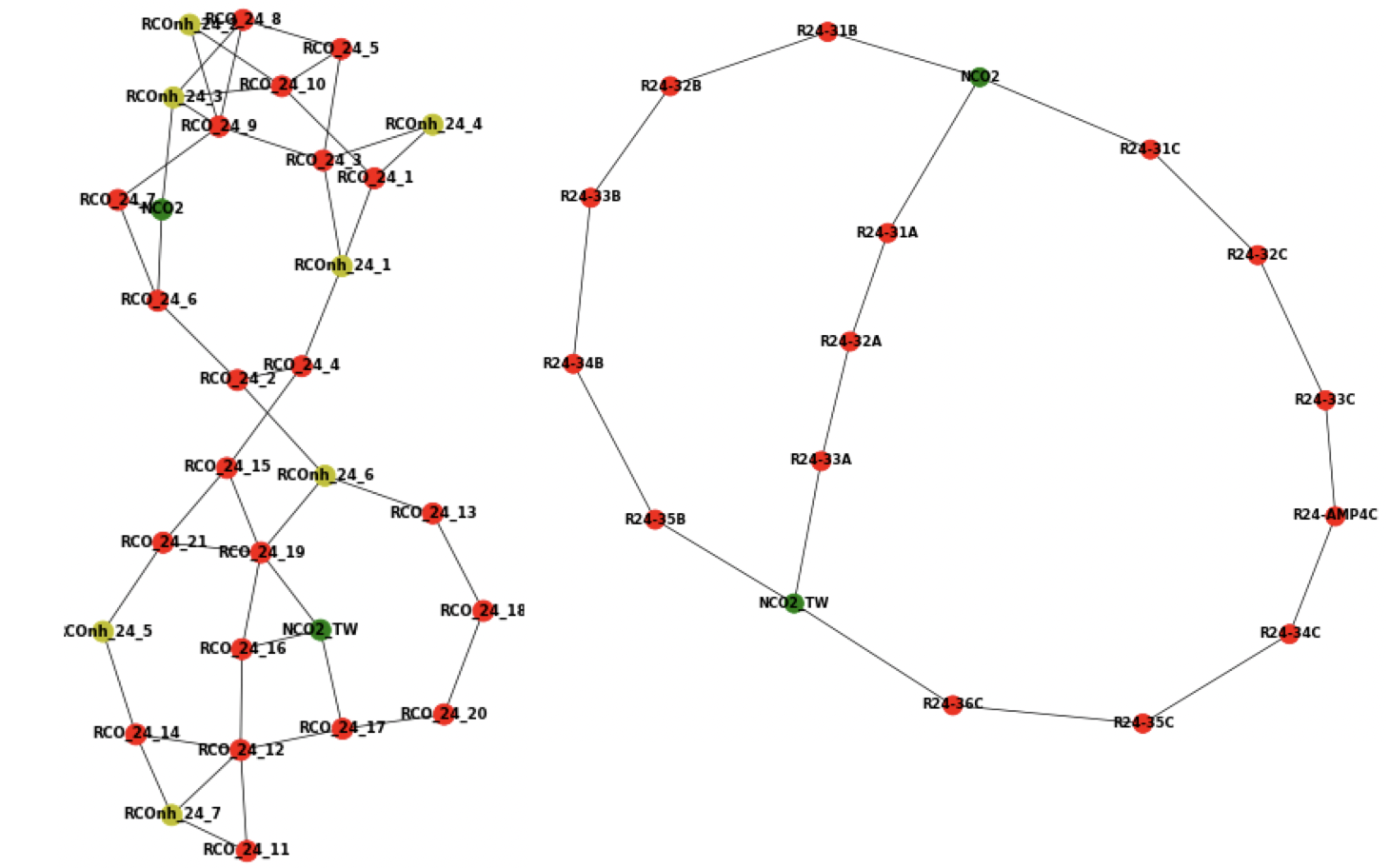}
  \caption{\small Example of metro core structures generated by MoleNetwork.}
  \label{MetroCoreExamples}
\end{figure}

An example of a metro mesh structure and a 3-ring-based structure generated by MoleNetwork are shown in Figure \ref{MetroCoreExamples}.

\begin{figure}[ht]
  \centering
  \includegraphics[width=200pt]{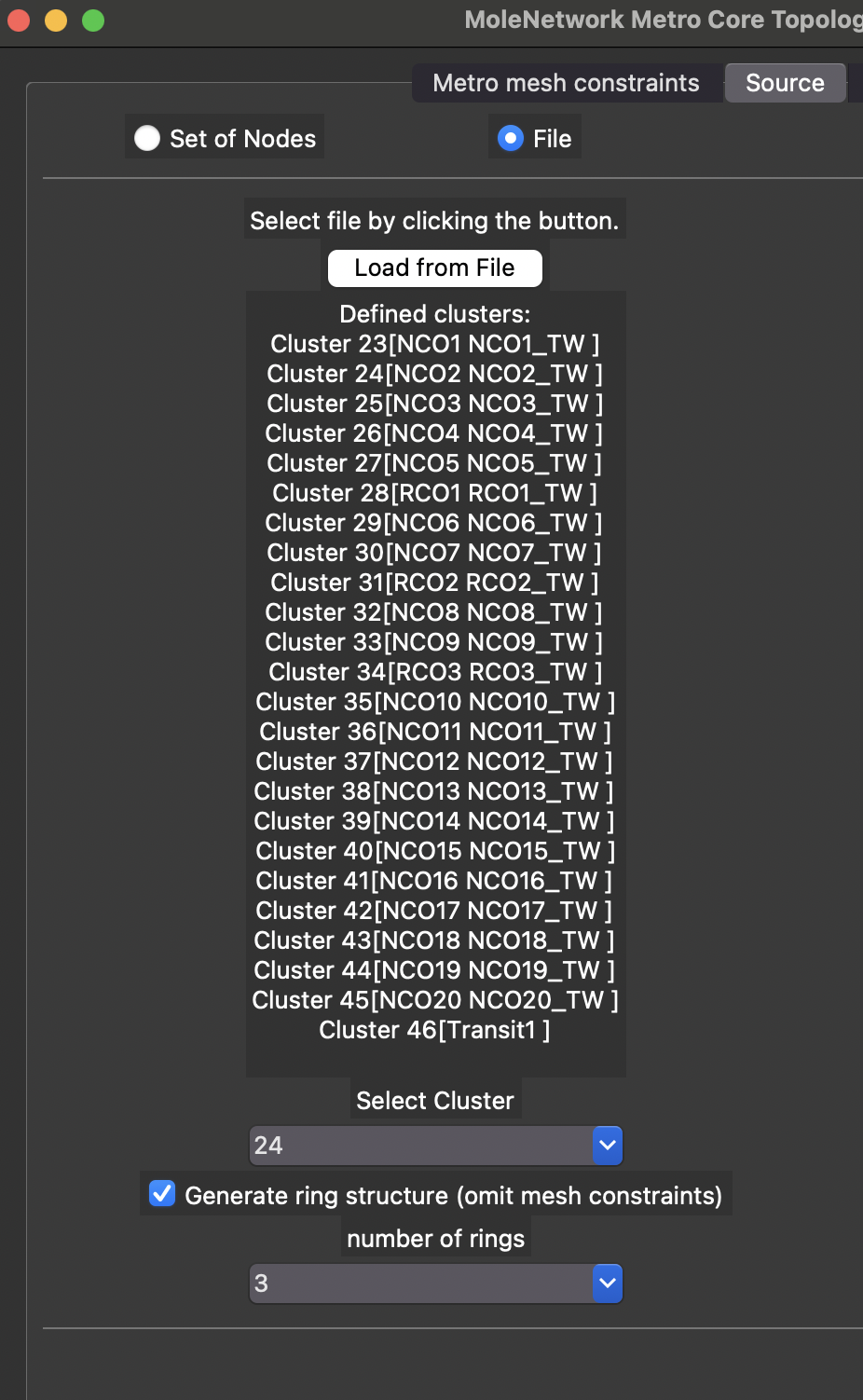}
  \caption{\small Reading backbone metro regions in the metro core GUI.}
  \label{MetroRegionsGUI}
\end{figure}

The generation of metro core networks also provides a graphical user interface that lets the user parameterize the system but also read the backbone generated Excel file, retrieve the clusters, and choose one of them to generate the metro core network (Figure \ref{MetroRegionsGUI}).

Additionally, if a single metro region is to be created, the GUI makes it possible to define the names of the national nodes instead of generating them sequentially.

Finally, the information can be stored in an Excel file (that can be the same as the backbone one merging some information) with the image stored as a file as well.

\subsection{Metro aggregation segment}
MoleNetwork also provides a way of generating metro aggregation networks mapping them to horseshoe/ring structures. The implementation uses the statistical information provided by Telecom Italia Mobile in the context of the Allegro project. However, as previously pointed out, a generic model is used so different implementations can be created.

\begin{figure}[ht]
  \centering
  \includegraphics[width=250pt]{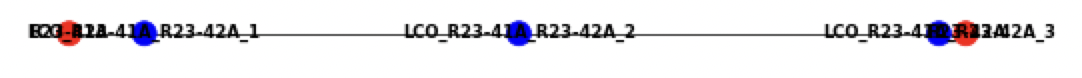}
  \caption{\small Example of a metro aggregation ring drawn as a straight line}
  \label{MetroAggHorseshoe}
\end{figure}

In this case, the structure is drawn as a straight line (Figure \ref{MetroAggHorseshoe}), with specific regional nodes used as ends. They are split into different segments corresponding to the hops, with specific lengths based on distance statistics. The actual shape of the structure would be, as previously commented, a ring or horseshoe, but in terms of representation, all the information is provided by the specific hop distances. 

The parameters provided to the algorithm for the generation of the horseshoe are the initial and final node names ($end1$ and $end2$), the initial index $idx$, and the number of hops for the horseshoe $hops$. Additionally, the user should provide a set of different possible length ranges for the total length of the horseshoe $len\_ranges$ and their corresponding probability of occurrence $perc_{length}$. Optional parameters include a prefix for the node names $pref$ and a mapping from node types to colors for the graphical representation $dict\_colors$.

The algorithm works as follows:
\begin{enumerate}
    \item One of the possible $len\_ranges$ (e.g. 10-20km) is randomly selected based on the probabilities provided $perc_{length}$. The total length $len\_horseshoe$ is retrieved using a uniform distribution to produce a value between both range ends.
   \item To define the location $loc\_local$ of the local COs (nodes in the horseshoe), $hops-1$ random values are generated between 1 and $len\_horseshoe$ and sorted in increasing order. These would be the specific distances between the origin and each of the nodes.
   \item A graph $horseshoe$ is created and the first end $end1$ is added to the graph. Its type is defined as "regional".
   \item The $hops-1$ nodes are added sequentially to the graph into the previously calculated distances $loc\_local$, generating edges between the previously inserted node and the new one. Their type is defined as "local" CO.
   \item The last end $end2$ is added to the graph and connected to the last of the generated nodes. Its type is defined as "regional" CO.
   \item Types of each node are mapped to colors using $dict\_colors$
   \item Positions are defined based on the $loc\_local$ values.
   \item A graphical representation of the created graph is prepared.
   \item For each node of the horseshoe, one of the end nodes ($end1$ or $end2$) is selected as the reference national node based on the distance, using the shortest path algorithm and a dummy node connected to the ends.
\end{enumerate}

\begin{figure}[ht]
  \centering
  \includegraphics[width=250pt]{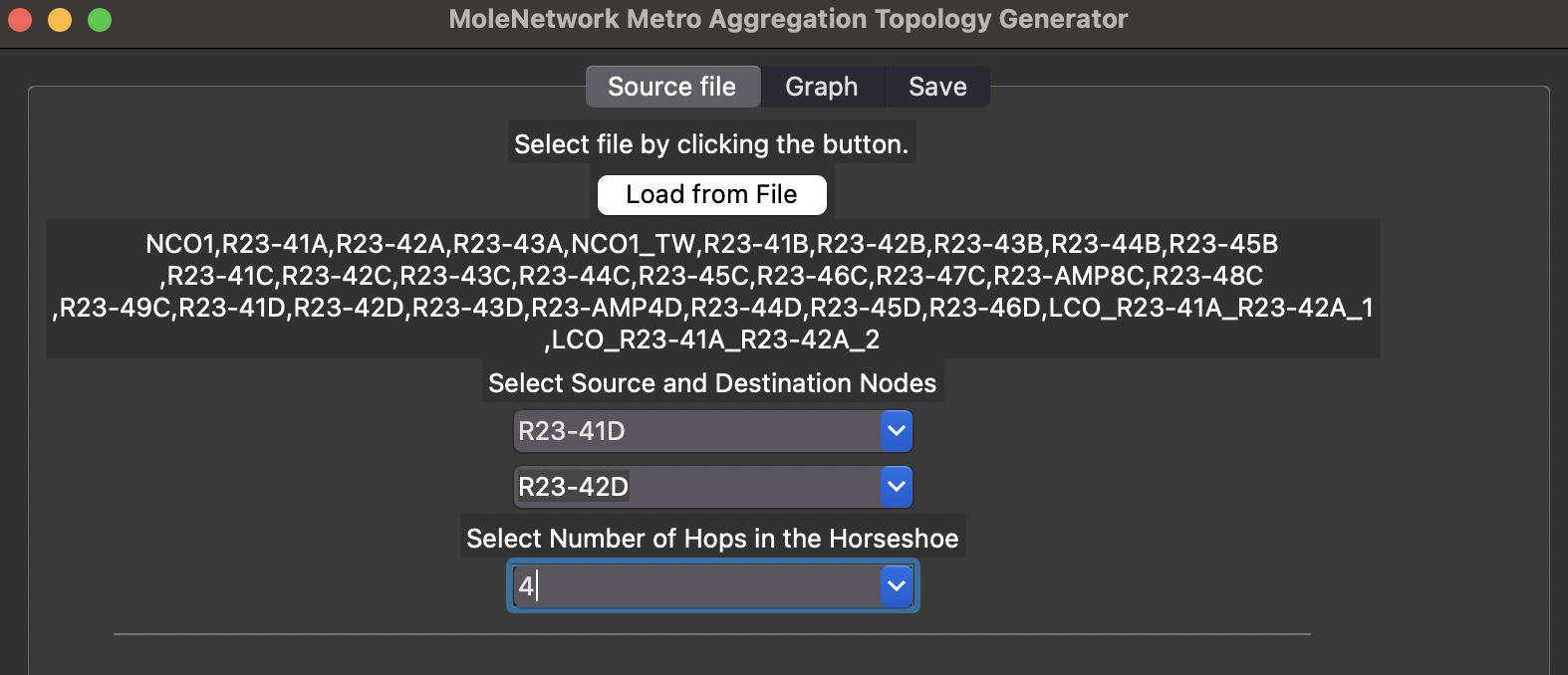}
  \caption{\small Selecting source and destination nodes in the Metro Aggregation GUI}
  \label{MetroAggGUI}
\end{figure}

A GUI is also available for metro aggregation generation (Figure \ref{MetroAggGUI}). In this case, the user can select an Excel file with a previously generated metro core network, load the information of that network, choose a source and destination node and the number of hops to generate the graph, and save the generated graph files.

\subsection{Backbone to metro aggregation integrated flow}

The supporting modules of the tool make it possible to create a flow to generate a complete topology from backbone to metro aggregation. 

\begin{figure}[ht]
  \centering
  \includegraphics[width=250pt]{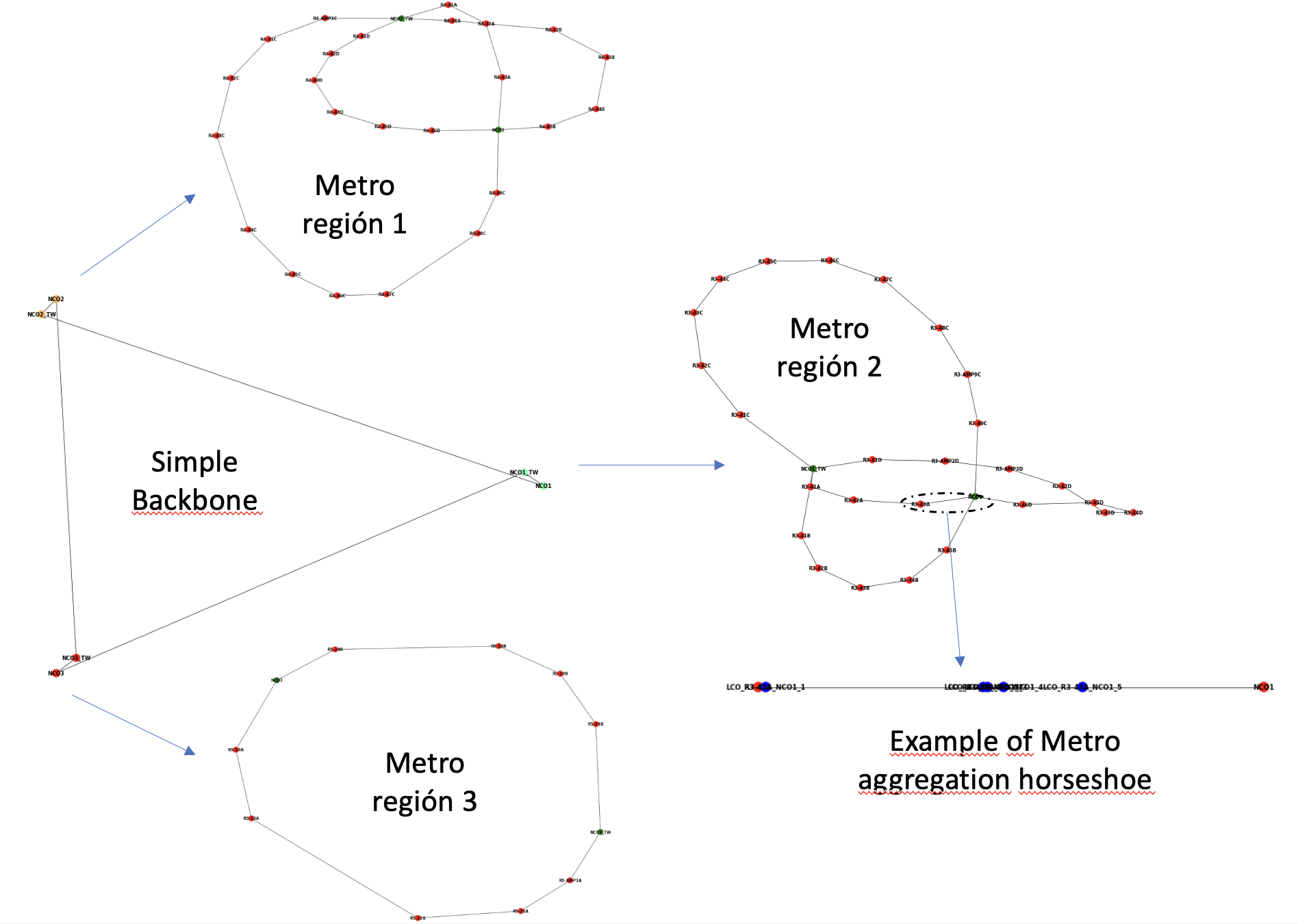}
  \caption{\small Example of the backbone to metro aggregation generation flow}
  \label{SampleGenerationFlow}
\end{figure}

An example with main parts of the topology is shown in Figure \ref{SampleGenerationFlow}. For simplicity, a small backbone network based on the twin approach is created with just six nodes which are automatically clustered together in three metro regions. 

Then, for each of the three clusters created, a ring structure is requested and generated. In this case, a 3-ring, a 4-ring, and a 2-ring structures are generated for those regions including nodes and amplifiers where required (based on statistical and distance information). 

In the created flow, we decided to build metro aggregation horseshoes for every two interconnected nodes (excluding amplifiers).

An example of one of the metro aggregation structures is shown in the right part of the figure.

All the information is integrated into a single Excel file for later processing or to be used in research or technical studies.

\begin{figure}[ht]
  \centering
  \includegraphics[width=250pt]{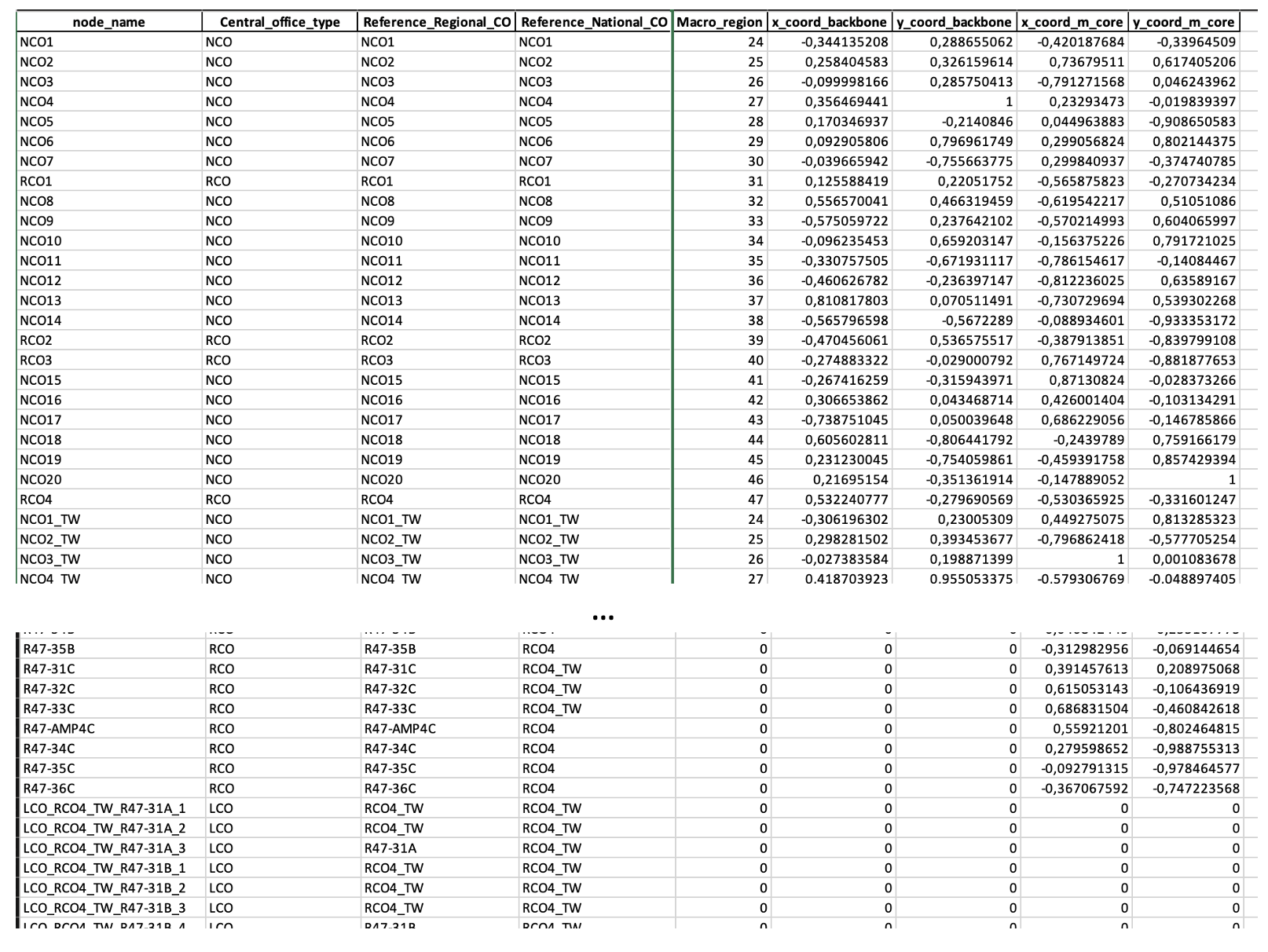}
  \caption{\small Example of the Excel output generated by the backbone to metro aggregation flow}
  \label{ExcelScreenshot}
\end{figure}

An example of the generated information from an Excel file for a backbone with 48 nodes and the related metro core and aggregation structures is shown in Figure \ref{ExcelScreenshot}

\section{Validation}
\label{validation}
Several experiments were carried out to generate networks using the tool and the statistics presented in Section \ref{statistics}.

To validate the tool, 1000 iterations were run for each of the three strategies presented in the backbone network. For the two first ones, the iterations were repeated for the different algorithms used for network layout.

As the degree distribution and the type generation are based on a random process using the defined proportions, and the proportions may not match an exact integer when applying them to a specific number of nodes, the result may not match exactly the provided values. Thus, the mean absolute percentage error (MAPE) metric has been used to compare the results with the information provided as an input parameter. Results when using the information from Table \ref{tab:topo_degree1} are shown in Table \ref{tab:degree_metric}. The comparison of the input degree distribution and the one with the best MAPE in each combination is shown in Table \ref{tab:best_degree_metric} (only showing two decimals). Notice that the results are, in general, very close to the input provided. The algorithm can be adapted to select the best result for a specific metric among a set of iterations. 

\begin{table}[ht]
\centering
\caption{Metrics for backbone degree distribution compared to input parameters}
    \begin{tabular}{|c|c|c|c|c|}
    \hline
        \multirow{2}{*}{Strategy} & \multirow{2}{*}{Algorithm} & \multicolumn{2}{c|}{{MAPE}} \\
    \cline{3-4}
         &  &
        Best & Average \\
    \hline
        Default & Spectral & 0.024 & 0.244\\
    \hline
        Default & Spring & 0.024 & 0.234\\ 
    \hline
        Default & Kamada & 0.024 & 0.242\\ 
    \hline
        Dual & Spectral & 0.041 & 0.376\\
    \hline
        Dual & Spring & 0.046 & 0.359\\ 
    \hline
        Dual & Kamada & 0.009 & 0.373\\ 
    \hline
        \multicolumn{2}{|c|}{{Pavan et al\cite{pavan}}} & 0.024 & 0.253\\ 
    \hline
    \end{tabular}
\label{tab:degree_metric}
\end{table}

\begin{table}[ht]
\centering
\caption{Comparison of inputs and degree proportions for best MAPE in the backbone scenario}
    \begin{tabular}{|c|c|c|c|c|c|c|}
    \hline
        Strategy & Algorithm & 2 & 3 & 4 & 5 & other\\

    \hline
        \multicolumn{2}{|c|}{{Inputs}} & 0.23 & 0.41 & 0.27 & 0.09 & 0\\
    \hline
        Default & Spectral & 0.23 & 0.42 & 0.27 & 0.08 & 0\\
    \hline
        Default & Spring & 0.23 & 0.42 & 0.27 & 0.08 & 0\\ 
    \hline
        Default & Kamada & 0.23 & 0.42 & 0.27 & 0.08 & 0\\ 
    \hline
        Dual & Spectral & 0.23 & 0.38 & 0.28 & 0.09 & 0.02\\
    \hline
        Dual & Spring & 0.22 & 0.43 & 0.26 & 0.09 & 0\\ 
    \hline
        Dual & Kamada & 0.23 & 0.41 & 0.27 & 0.09 & 0\\ 
    \hline
        \multicolumn{2}{|c|}{{Pavan et al\cite{pavan}}} & 0.23 & 0.42 & 0.27 & 0.08 & 0 \\ 
    \hline
    \end{tabular}
\label{tab:best_degree_metric}
\end{table}

Although the algorithms are not originally created to match specific distance ranges and weights as the ones provided in Table \ref{tab:topo_link1}, it is possible to apply a factor (configurable to reduce an error metric) to approximate the generated topology distances to the ones required by the user. Additionally, a number of executions can be run to select the model that approximates better the required ranges of distances. Thus, we have tested how the different models fit the specific ranges provided in Table \ref{tab:topo_link1} selecting the best result out of the 1000 executions. The result is shown in Table \ref{tab:best_distance_metric}. It looks like the best approximation for this set of distance ranges and proportions is the use of the Spectral algorithm. This may change with different distance distributions.

\begin{table}[ht]
\centering
\caption{Comparison of backbone distance range percentages for the best match with the provided ranges}
    \begin{tabular}{|c|c|c|c|c|c|c|}
    \hline
        Strat. & Algo. & 0-50 & 50-100 & 100-200 & 200-400 & 400-600\\

    \hline
        \multicolumn{2}{|c|}{{Inputs}} & 15.5 & 16.9 & 33.8 & 25.4 & 8.5\\
    \hline
        Default & Spectral & 12.7 & 17.7 & 34.1 & 26.6 & 8.9\\
    \hline
        Default & Spring & 1.2 & 2.5 & 21.0 & 66.7 & 8.6\\ 
    \hline
        Default & Kamada & 0 & 0 & 16.7 & 74.3 & 9.0\\ 
    \hline
        Dual & Spectral & 16.4 & 22.4 & 28.3 & 25.4 & 7.5\\
    \hline
        Dual & Spring & 14.5 & 21.7 & 18.8 & 34.8 & 10.2\\ 
    \hline
        Dual & Kamada & 15.6 & 15.6 & 16.9 & 42.8 & 9.1\\ 
    \hline
        \multicolumn{2}{|c|}{{Pavan et al\cite{pavan}}} & 0 & 16.5 & 47.1 & 28.2 & 8.2\\ 
    \hline
    \end{tabular}
\label{tab:best_distance_metric}
\end{table}

In the case of the Metro Mesh generation, we used the same degree distribution as input for 1000 runs of the experiment with the three different algorithms. We decided to use 95 nodes with 5 national central offices in the experiment as in the example in Figure \ref{MetroMesh_Sample}. Results are shown in Tables \ref{tab:degree_metric_mesh} and \ref{tab:best_degree_metric_mesh}. 

Again, they have been approximated to the link distance distribution, in this case from Table \ref{tab:topo_metro_link1}. Values for the best outputs in the 1000 iterations are shown in \ref{tab:best_distance_metric_mesh}.

\begin{table}[ht]
\centering
\caption{Metrics for metro mesh degree distribution compared to input parameters}
    \begin{tabular}{|c|c|c|c|c|}
    \hline
        \multirow{2}{*}{Strategy} & \multirow{2}{*}{Algorithm} & \multicolumn{2}{c|}{{MAPE}} \\
    \cline{3-4}
         &  &
        Best & Average \\
    \hline
        Default & Spectral & 0.135 & 0.408\\
    \hline
        Default & Spring & 0.131 & 0.415\\ 
    \hline
        Default & Kamada & 0.095 & 0.414\\ 
    \hline
    \end{tabular}
\label{tab:degree_metric_mesh}
\end{table}

\begin{table}[ht]
\centering
\caption{Comparison of inputs and degree proportions for best MAPE in the metro scenario}
    \begin{tabular}{|c|c|c|c|c|c|c|}
    \hline
        Strategy & Algorithm & 2 & 3 & 4 & 5 & other\\

    \hline
        \multicolumn{2}{|c|}{{Inputs}} & 0.23 & 0.41 & 0.27 & 0.09 & 0\\
    \hline
        Default & Spectral & 0.19 & 0.48 & 0.26 & 0.07 & 0\\
    \hline
        Default & Spring & 0.21 & 0.47 & 0.24 & 0.08 & 0\\ 
    \hline
        Default & Kamada & 0.21 & 0.46 & 0.24 & 0.09 & 0\\ 
    \hline
    \end{tabular}
\label{tab:best_degree_metric_mesh}
\end{table}

\begin{table}[ht]
\centering
\caption{Comparison of metro mesh distance range percentages for the best match with the provided ranges}
    \begin{tabular}{|c|c|c|c|c|c|c|}
    \hline
        Strat. & Algo. & 0-10 & 10-40 & 40-80 & 80-120 \\

    \hline
        \multicolumn{2}{|c|}{{Inputs}} & 39 & 37 & 21 & 3 \\
    \hline
        Default & Spectral & 36 & 40.7 & 20 & 3.33 \\
    \hline
        Default & Spring & 21 & 47.4 & 24.2 & 7.4 \\ 
    \hline
        Default & Kamada & 0 & 2.5 & 78.9 & 18.6\\ 
    \hline
    \end{tabular}
\label{tab:best_distance_metric_mesh}
\end{table}

Validation was completed with the rest of the metro core and aggregation structures by checking that they were created following the statistics from Table \ref{tab:topo_metro_ring1} for the metro N-rings and Table \ref{tab:topo_agg_ring1} for the metro aggregation horseshoes. 

\section{Conclusion}
\label{conclusion}
This paper has described an open-source tool for developing backbone, metro core and metro aggregation networks based on statistical parameters that can be customized by the user. It provides several graphical user interfaces to make the generation process easy and it also enables the creation of a flow to define the complete backbone of metro aggregation segments of a particular topology. 

The different algorithms used for generating the various network segments have been detailed and the complete generation flow has also been described showing an example with a small network.

Finally, validation was completed to check that the degree distribution was followed and to show that it is possible to approximate the results to a distribution of distances for the links to simulate a similar network.

Future work includes the incorporation of the access segment and the definition of new strategies for topology generation based on Artificial Intelligence (e.g. Machine Learning) among other topics.

\section*{Acknowledgment}
The authors would like to acknowledge the support of EU-funded ALLEGRO projects (grant No.101092766) and R\&D project PID2022-136684OB-C21 (Fun4Date) funded by the Spanish Ministry of Science and Innovation MCIN/AEI/ 10.13039/501100011033

\bibliographystyle{unsrt}  
\bibliography{MoleNetwork}

\end{document}